\documentclass{llncs}
\usepackage{graphicx}
\usepackage{subfigure}
\usepackage{amssymb}
\usepackage{layouts} %to show linewidth
\newcommand{\V}[1]{\ensuremath{\vec{#1}}} %vector
\newcommand{\M}[1]{\ensuremath{\mathsf{#1}}}

\newcommand{\Sc}[1]{\ensuremath{#1}} %scalar
\newcommand{\Sd}[2]{\ensuremath{\Sc{#1}_{#2}}} %scalar with subscript
\newcommand{\Su}[2]{\ensuremath{\Sc{#1}^{#2}}} %scalar with superscript

\newcommand{\Md}[2]{\ensuremath{\M{#1}_{#2}}} %matrix with subscript
\newcommand{\Mu}[2]{\ensuremath{\M{#1}^{#2}}} %matrix with superscript
\newcommand{\eg}{e.g.~} %e.g.
\newcommand{\ie}{i.e.~} %i.e.

\newcommand{\transpose}{T} %transpose

\newcommand{\pinv}[1]{{#1}^\mathrm{+}}
\newcommand{\nullS}[1]{\ensuremath{\Md{N}{#1}}}

\begin{document}
	\title{Advanced Symbolic Time Series Analysis\\ in Cyber Physical Systems}
%TODO only ASTSA or ASTSA-CPS?
%\author{Roland Ritt\inst{1} \and Paul O'Leary \inst{1} \and Matthew Harker  \inst{1}}
\author{Roland Ritt \and Paul O'Leary \and Christopher Josef Rothschedl \and Matthew Harker}
\institute{Institute for Automation, University of Leoben, Leoben, Austria \email{roland.ritt@unileoben.ac.at}}
\maketitle
 \newline

%
%
%\begin{abstract}
%	test Abstract
%\end{abstract} %Title, authors and Abstract
	This paper presents advanced symbolic time series analysis (ASTSA) for large data sets emanating from cyber physical systems (CPS).
The definition of CPS most pertinent to this paper is:
\emph{
	A CPS is a system with a coupling of the cyber aspects of computing and communications with the physical aspects of dynamics and engineering that must abide by the laws of physics.
	This includes sensor networks, real-time and hybrid systems} \cite{baheti2011cyber}.
To ensure that the computation results conform to the laws of physics a linear differential operator (LDO) is embedded in the processing channel for each sensor.
In this manner the dynamics of the system can be incorporated prior to performing symbolic analysis. A non-linear quantization is used for the intervals corresponding to the symbols.
The intervals are based on observed modes of the system, which can be determined either during an exploratory phase or on--line during operation of the system.
A complete processing channel (see Fig.~\ref{fig:SCLA})  is called a single channel lexical analyser; one is made available for each sensor on the machine being observed.

The implementation of LDO in the system is particularly important since it enables the establishment of a causal link between the observations of the dynamic system and their cause.
Without causality there can be no semantics and without semantics no knowledge acquisition based on the physical background of the system being observed.
Correlation alone is not a guarantee for causality\footnote{
	Consider an exothermic system with a high activation energy.
	We must include the exothermic model if we are to establish causality, correlation alone will lead to erroneous interpretation.}

\begin{figure}
	\centering
	\begin{subfigure}{
		\includegraphics[width=0.31\textwidth]{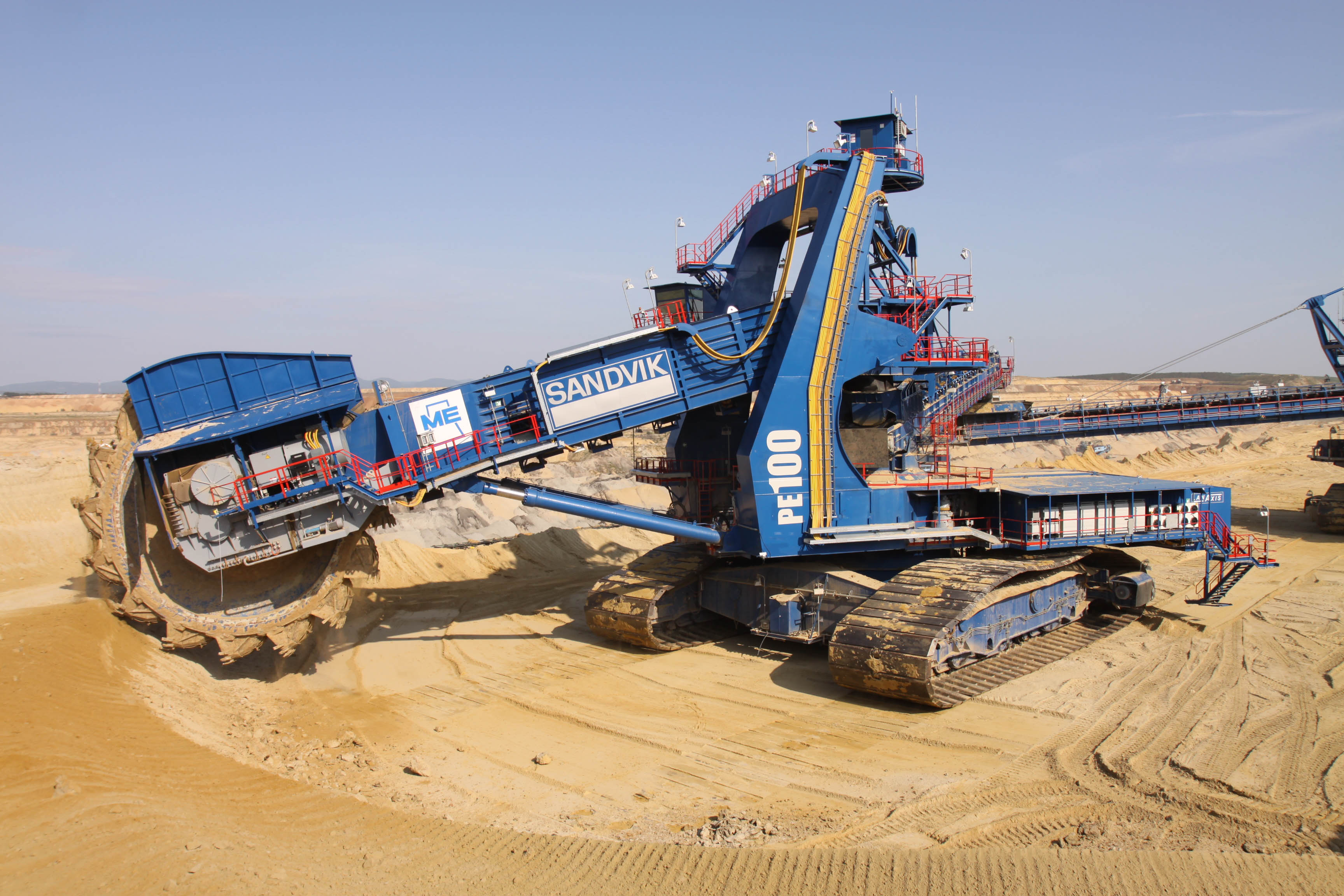}}
	\end{subfigure}
\hfill
	\begin{subfigure}{
		\includegraphics[width=0.31\textwidth]{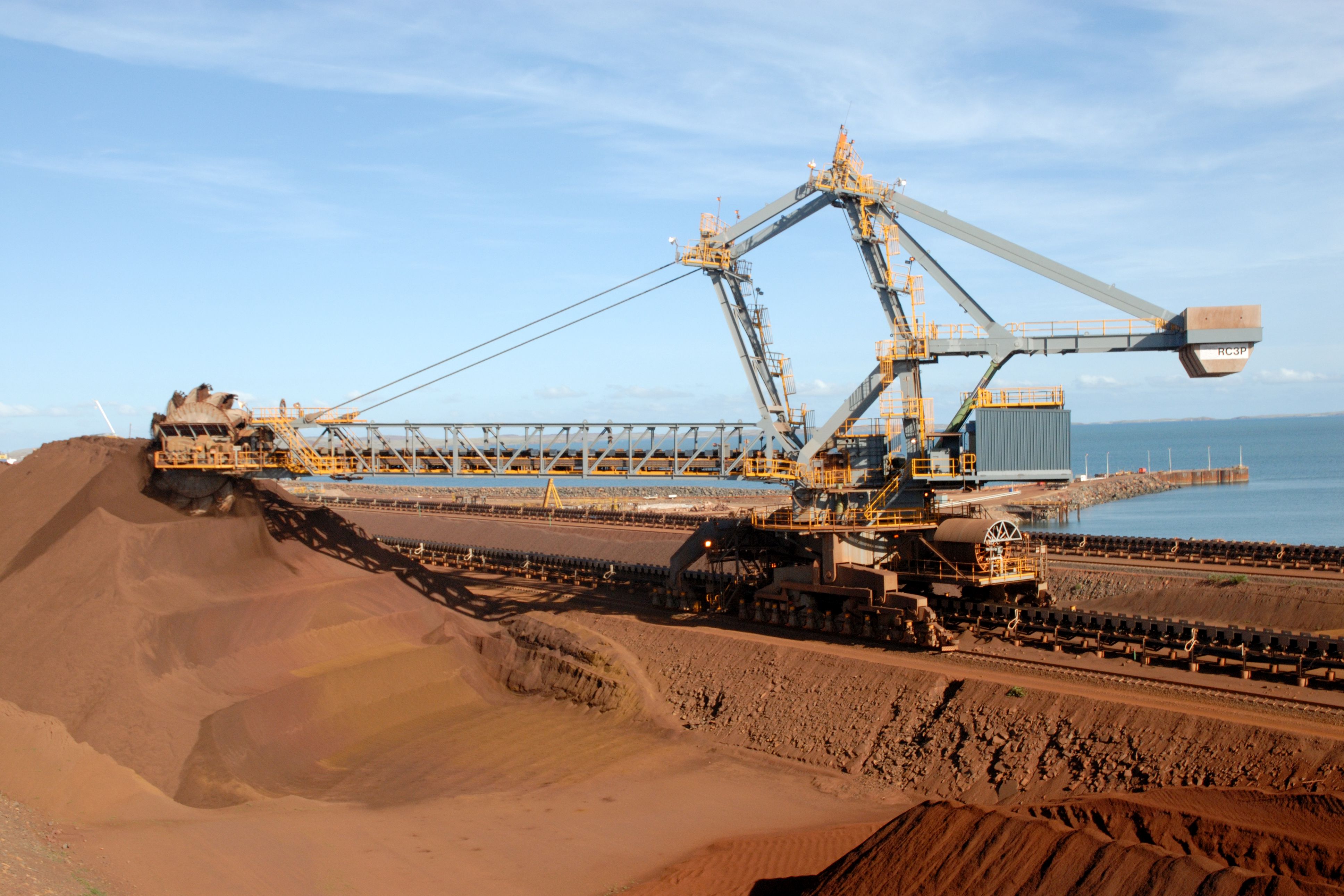}}
	\end{subfigure}
\hfill
	\begin{subfigure}{
		\includegraphics[trim={0.5cm 0cm 2.3cm 0cm}, clip,width=0.31\textwidth]{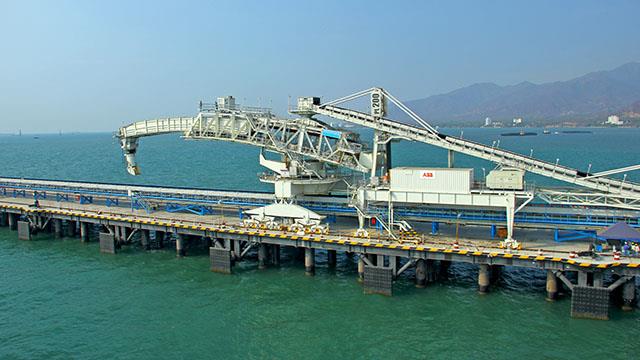}}
	\end{subfigure}
\caption{Examples of machines to which the analysis is applied. Image courtesy: Sandvik}
\label{fig:Machines}
\end{figure}

This work was originally motivated from the observation of large bulk material handling systems, see Fig.~\ref{fig:Machines} for three examples of such systems.
Typically, there are $n= 150 \ldots 250$ sensors per machine, and data is collected in a multi rate manner; whereby general sensors are sampled with $f_s = 1 \, Hz$ and vibration data being sampled in the kilo-hertz range.
	% 20.06.2017
%
\section{Local Linear Differential Operators (LDO)}
%Collecting (big) data is an important topic in a broad field of applications.
%%
%Data availability although not knowing what to do with that data is getting important in nearly every industry.
%%
%Cloud and distributed computation services allow to easy handle and analyse such big data.
%% TODO cite
%%
%A common system premise of such distributed computations is the map-reduce architecture.
%% TODO cite
%%
%Therefore the data are partitioned and local computations are performed (map) before combining the results of these computations (reduce).
%%
%Since we deal with time series in cyber physical systems new techniques for local computations including the physics of the system (described by differential equations) have to be developed.
%%
Although processing the entire 'large' time series is a common practice in exploratory data analysis, reliable local computations (implemented as streaming algorithms) are preferred in on--line data processing.
Since in this work we deal with time series emanating from cyber physical systems new techniques for local computations including the physics of the system (described by differential equations) have to be developed.

An ordinary differential equation (ODE) of the form
\begin{equation}
\Sd{a}{d}\left(t\right) \Su{y}{\left(d\right)} \left(t\right)+ \Sd{a}{d-1} \left(t\right) \Su{y}{\left(d-1\right)} \left(t\right) + \dots + \Sd{a}{0} \left(t\right)\Su{y}{\left(0\right)} \left(t\right) = \Sc{g}\left(t\right)
\label{equ:ODE}
\end{equation}
can be described using a linear differential operator (LDO) $\M{D}$ \cite{Lanczos1961} such that $\Mu{D}{(i)} \Sc{y}\left(t\right) = \Su{y}{\left(i\right)}\left(t\right)$ where $\Sc{y}$ is a function of $\Sc{t}$, $\Su{y}{\left(i\right)}$ is the $\Sc{n}$-th derivative with respect to $\Sc{t}$ and $\Sc{g}\left(t\right)$ is the exciting function, in our case the noisy sensor data.
This yields to the notation \cite{OLeary2015Inverse}
\begin{equation}
	\Sd{a}{d}\left(t\right) \Mu{D}{(d)} \Su{y}{\left(d\right)} \left(t\right)+ \Sd{a}{d-1} \left(t\right)  \Mu{D}{(d-1)} \Su{y}{\left(d-1\right)} \left(t\right) + \dots + \Sd{a}{0} \left(t\right) \Mu{D}{(0)} \Su{y}{\left(0\right)} \left(t\right) = \Sc{g}\left(t\right).
\end{equation}
Factoring $\Sc{y}\left(t\right)$ leads to the compact formulation of the model
\begin{equation}
\M{L} y \left(t\right) = \Sc{g}\left(t\right),
\label{equ:LDOgeneral}
\end{equation}
with
\begin{equation}
	\M{L} \triangleq 	\Sd{a}{d}\left(t\right) \Mu{D}{(d)} + \Sd{a}{d-1} \left(t\right)  \Mu{D}{(d-1)}  + \dots + \Sd{a}{0} \left(t\right) \Mu{D}{(0)}.
\end{equation}

In the discrete case (\ref{equ:LDOgeneral}) can be formulated as matrix equation.
Solving this equation for $\Sc{y}$ is an inverse problem which can be solved numerically in a discrete sense by
\begin{equation}
	\V{y} = \pinv{\M{L}} \V{g} + \nullS{\M{L}}\V{\alpha},
	\label{equ:SolLDO}
\end{equation}
where $\V{y}$ is the solution to the inverse problem, $\pinv{\M{L}}$ is the pseudo-inverse of $\M{L}$,  $\nullS{\M{L}}$ is an orthonormal basis function set of the null space of $\M{L}$ , $\V{\alpha}$ is a coefficient vector for the null space (computed by initial- and/or the boundary-values) and $\V{g}$ is the noisy time series data vector.
Algebraic implementations for the solution of such problems can be found in \cite{Gugg2015,harker2013DOP,Gugg2015phd,OLeary2008Algebraic}.
%
%
%Gugg~et~al. \cite{Gugg2015} presented a mathematical framework for an efficient solution of such inverse problems including arbitrary constraints.
%
%%
%Note: Various basis function sets are suitable for setting up the derivative matrices $\Mu{D}{(i)}$ \cite{Gugg2015phd}.
%%
%In this work discrete orthogonal polynomials \cite{OLeary2008Algebraic} used due to their numerical behaviour (\MATLAB-code is available  for generating them \cite{harker2013DOP}).
%%
%

The LDO, and their inverses, can be implemented as local operators and efficiently computed using a convolutional approach.
This is basically a streaming-algorithm and thus suitable for big-data processing.

Furthermore, the covariance of the solution (\ref{equ:SolLDO}) is simply propagated as
\begin{equation}
	\Md{\Lambda}{y} = \pinv{\M{L}}  \Md{\Lambda}{g}  \left(\pinv{\M{L}}\right)^\transpose.
	\label{equ:covSolution}
\end{equation}
Using $\Md{\Lambda}{y}$ as an estimate for the covariance in conjunction with the student-$t$ and/or F-distribution permits the estimation of a confidence interval over the complete solution and allows the computation of a prediction interval for future values.

That is, the approach presented here to implementing linear differential operators not only permits the solution of embedded system dynamics but also yields a confidence interval for the
predicted values of the dynamics.
%
%
%\textbf{Alt}
%Although the given formulation enables solving the problem considering each point in the given time-series (which in fact leads to large matrices and thus to heavy computations) a local view on the data partitions the problem into small sub-problems.
%%
%Therefore, the problem is formulated and solved only on for small portion of the data (window with length $\Su{l}{s}$).
%%
%If the sampling rate stays constant (equally spaced timestamps), the solution of the local problem for each point can be calculated via a convolutional approach by convolving the centre line of $\M{L}$  with the data and correcting only the end points.
%%
%This is basically a streaming-algorithm and thus suitable for big-data processing. 
	%\input{Sec_LocalErrorCalc}
	% 20.06.2017
%
\section{Symbolic Time Series Analysis}
The availability of the sensor signals, their regularized derivative and/or the application of a LDO permits the implementation of an advanced symbolic time series analysis (ASTSA) which includes the modelling of the system dynamics.
As a result the time series~(TS) can be discretized and compressed using unique symbols for different intervals (the so called alphabet).
This step is named lexical analysis.
%
%TODO how timeseries spelled
%
A number of methods for the selection of the symbol intervals based on, \eg, equal probability, variance or entropy can be found in literature \cite{Lin,Veenman2002,Chau1999a,Keogh2002a,Daw2003}.
Here, in a new approach, we define the intervals to correspond to the modes of the dynamic system in operation, \ie each symbol corresponds to a mode or portion of a mode which should be identified.
Commonly controllers are designed to operate optimally in a number of specific but distinct modes of the dynamic system.

\begin{figure}
	\centering
	\includegraphics[trim={2cm 5.7cm 1cm 0.5cm}, clip, width=1\textwidth]{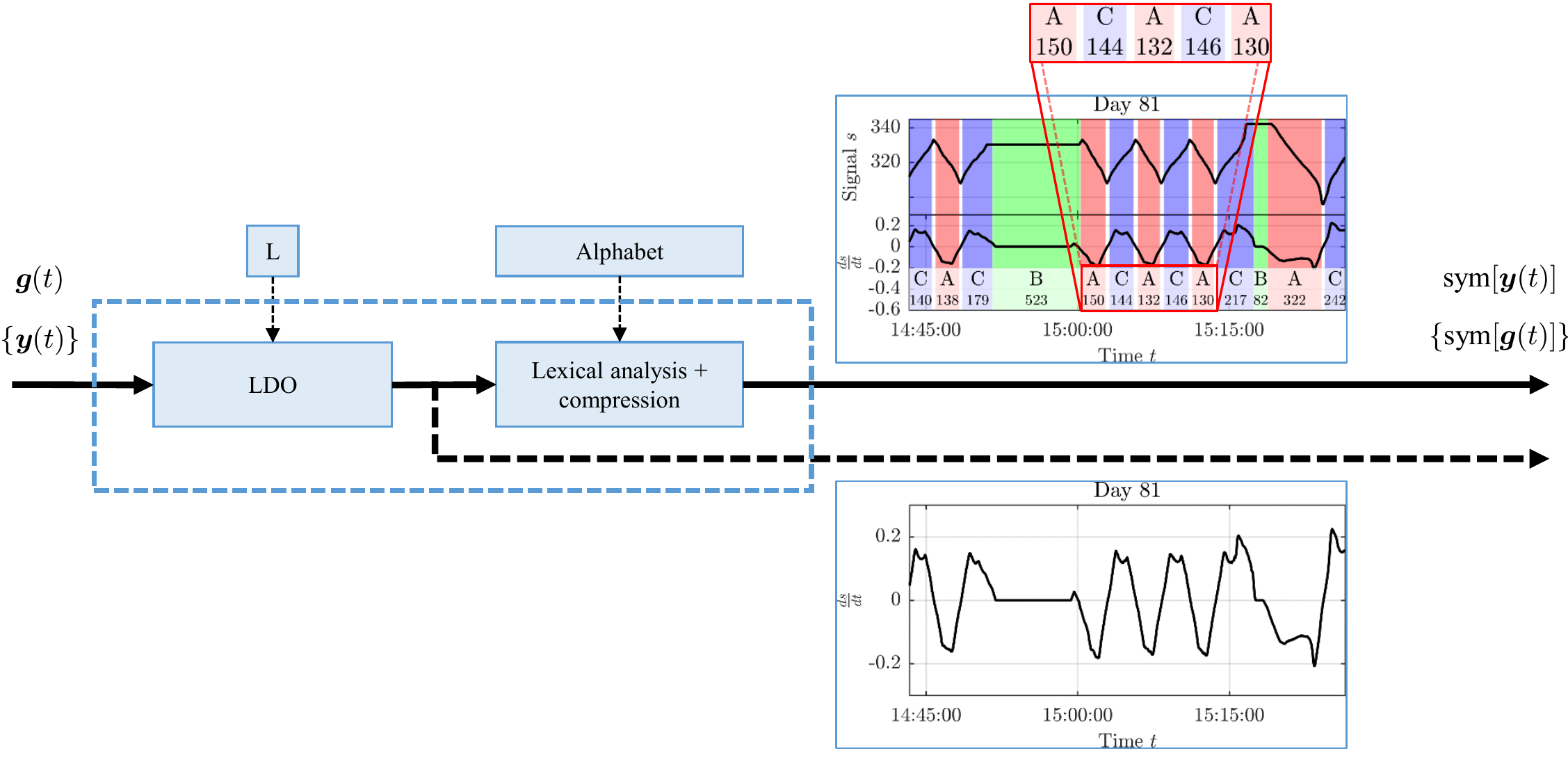}
	\label{fig:SCLA}
	\caption{A single channel lexical analyser (SCLA)}
\end{figure}

\begin{figure}[h]
	\centering
	\includegraphics[width=1\textwidth]{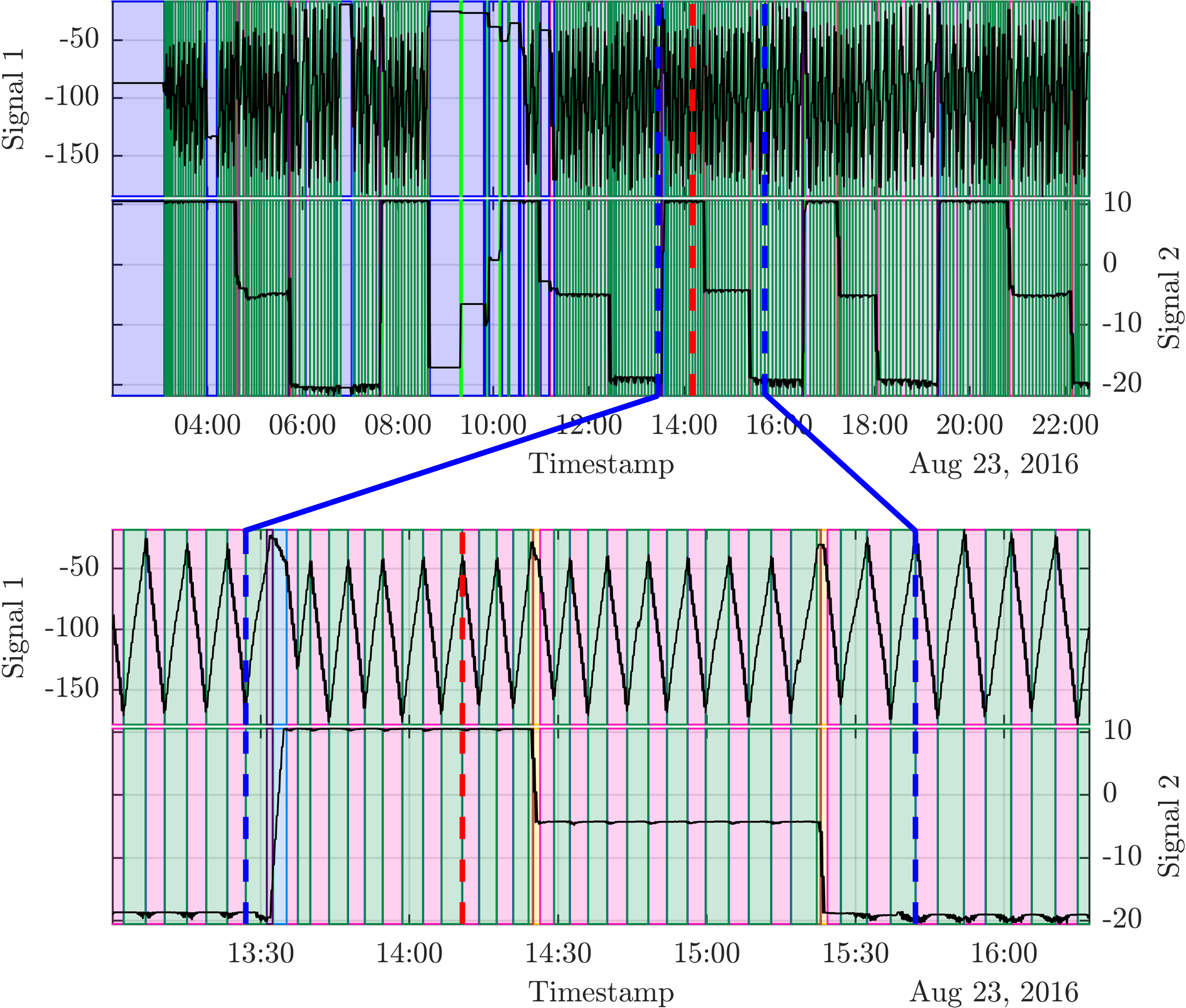}
	\caption{Operation mode 1; the coloured areas illustrate the output of the MCLA; different colours represent different combinations of symbols from the SCLA of each channel (in this case two channels); the alphabet used for signal 1 and 2 consists of the three symbols [\texttt{u}, \texttt{s}, \texttt{d}]. Top: machine working in operation mode 1 with longer interrupts  in-between (light blue area - both signals are stationary); Bottom: snippet of the signal showing the typical repeating pattern of operation mode 1.}
	\label{fig:MCLA1}
\end{figure}

\begin{figure}[h]
	\centering
	\includegraphics[width=1\textwidth]{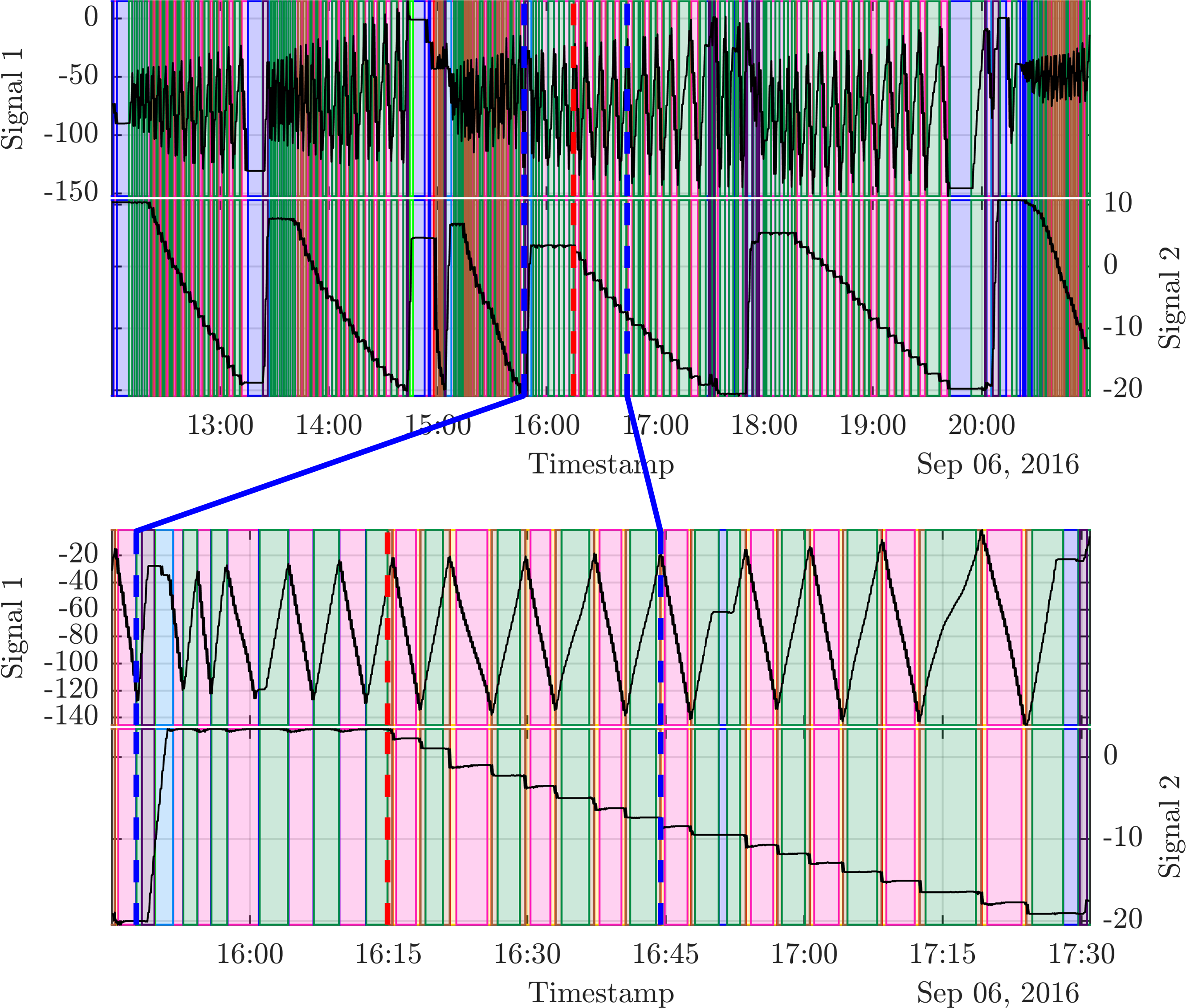}
	\caption{Operation mode 2; the coloured areas illustrate the output of the MCLA; different colours represent different combinations of symbols from the SCLA of each channel (in this case two channels); the alphabet used for signal 1 and 2 consists of the three symbols [\texttt{u}, \texttt{s}, \texttt{d}]. Top: machine working in operation mode 2 with  interrupts in-between (light blue area - both signals are stationary); Bottom: snippet of the signal showing the typical repeating pattern of operation mode 2.}
	\label{fig:MCLA2}
\end{figure}

In a next step, connected sequences with the same symbol can be compressed to a single symbol predicated with its length.
The combination of applying a LDO, lexical analysis of the derived signal and compression is called single channel lexical analyser~(SCLA), see Fig.~\ref{fig:SCLA}.
Combining the output of multiple SCLA is called multi channel lexical analyser~(MCLA).
Two examples of symbolic time series analysis using MCLA are demonstrated in Fig.~\ref{fig:MCLA1} and Fig.~\ref{fig:MCLA2}).
For signal 1 and signal 2 the alphabet consists of the three symbols [\texttt{u}, \texttt{s}, \texttt{d}] assigned to the direction of the signal (up, stationary, down).
The figures show two operation modes from the same machine.
It can be clearly seen, that the operation modes of the machine have a different symbolic representation (visualized as different shaded colours in the plots) and allow a fast intuitive inspection and characterization of the signal.
The signal range from the first dashed-blue line to the dashed-red line (marked in both plots) have the same symbolic representation in both modes, whereas the portion of the signal after the dashed-red line shows a different colour-code for each mode.
%\begin{figure}
%	\centering
%	\begin{subfigure}[Operation mode 1; the symbol representation of the signals between the blue dashed lines using MCLA is: ?????]{
%			\includegraphics[width=1\textwidth]{figures/Mode1gesamt.png}}
%	\end{subfigure}
%	\vfill
%	\begin{subfigure}[Operation mode 2; the symbol representation of the signals between the blue dashed lines  using MCLA is: ??]{
%			\includegraphics[width=1\textwidth]{figures/Mode2gesamt.png}}
%	\end{subfigure}
%	\caption{Two examples of symbolic time series analysis using MCLA; the different shaded colors represent different combinations of symbols from the SCLA; the alphabet used for signal 1 and 2 consists of the three symbols [\texttt{u}, \texttt{O}, \texttt{d}] assigned to the direction of the signal (up, no change, down); the combination of symbols generated by the SCLA of each signal is presented in the captions above. The first signal protions have the same shape - represented as the same symbol sequence (line 1), after the red dashed line the signals differ (different operation modes) - this can be clearly seen in the second and third symbol sequence lines }
%	\label{fig:MCLA}
%\end{figure}
%
%
%

\begin{figure}
	\centering
	\begin{subfigure}[Operation mode 1]{
			\includegraphics[scale=1]{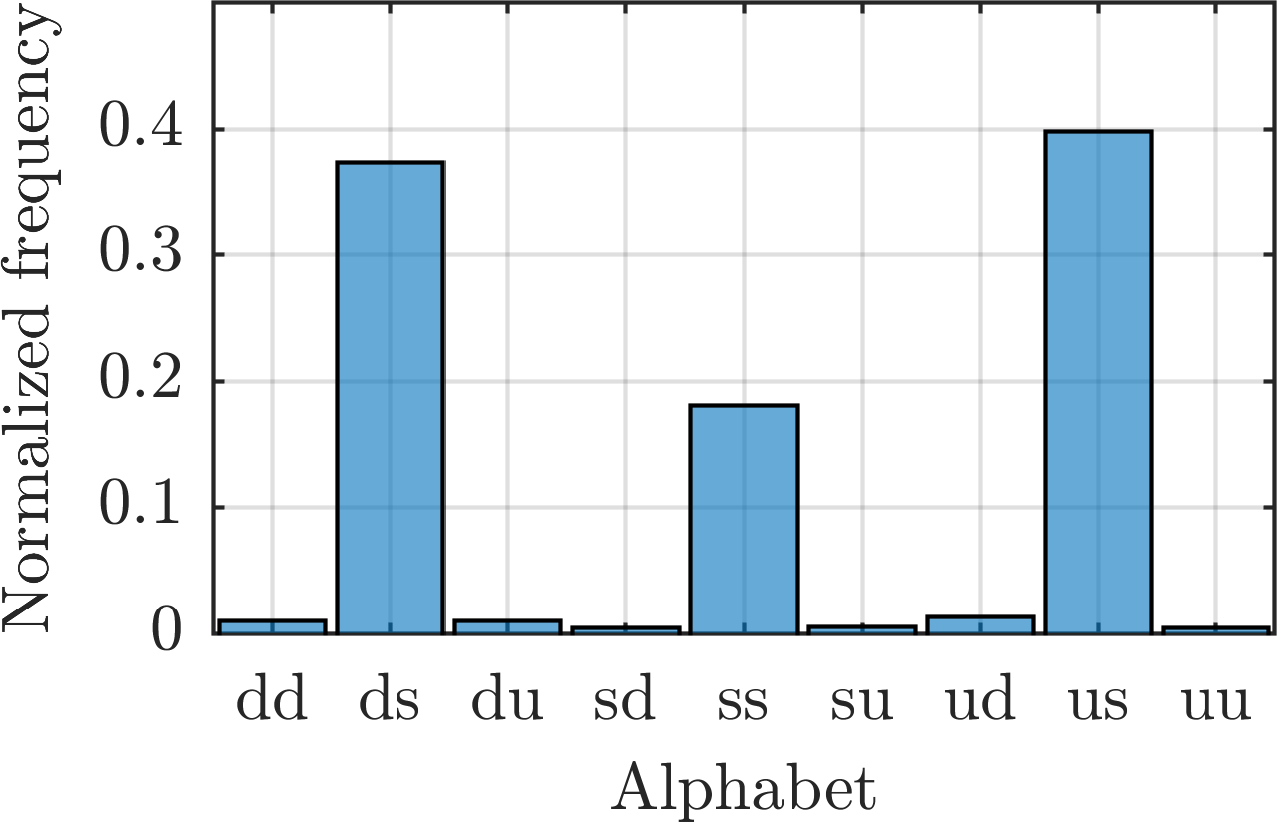}}
	\end{subfigure}
	\hfill
	\begin{subfigure}[Operation mode 2]{
			\includegraphics[scale=1]{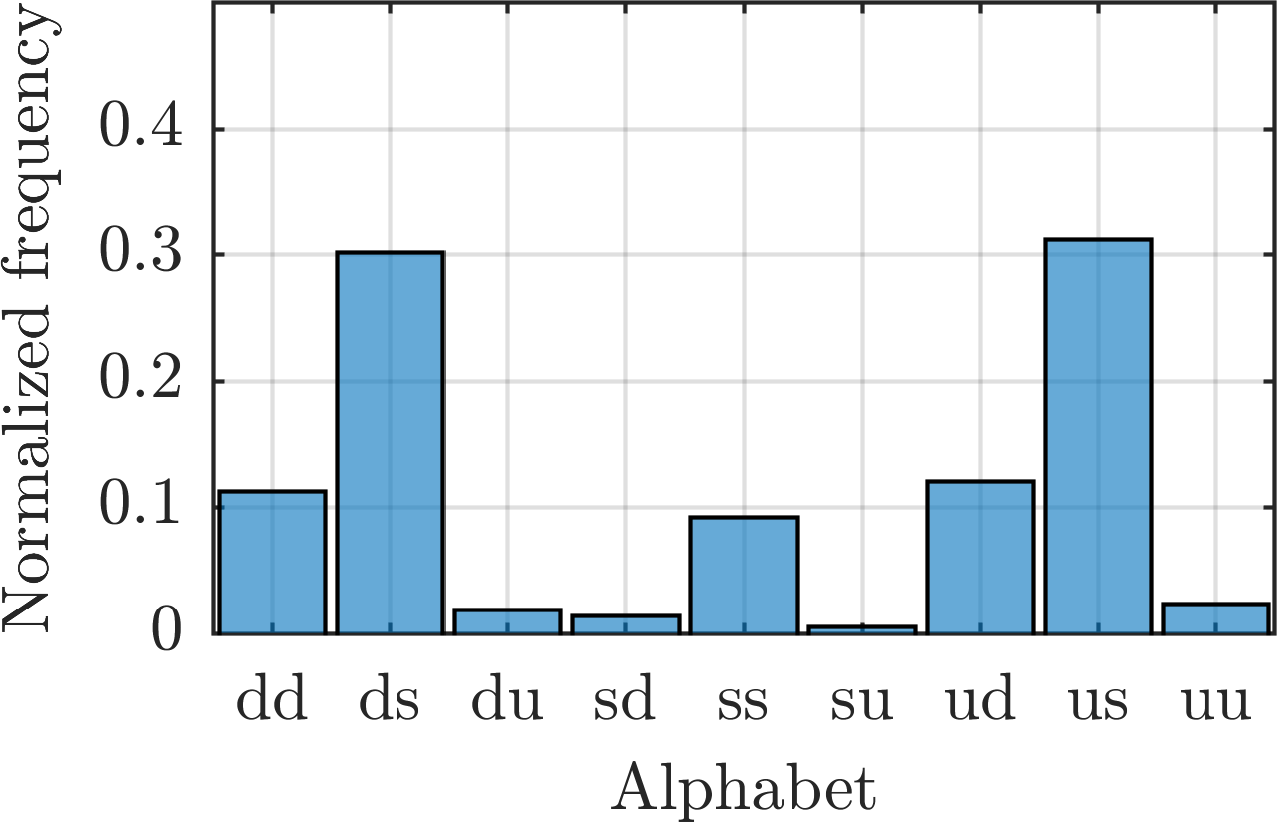}}
	\end{subfigure} \\
		\begin{subfigure}[Operation mode 1 - snippet]{
			\includegraphics[scale=1]{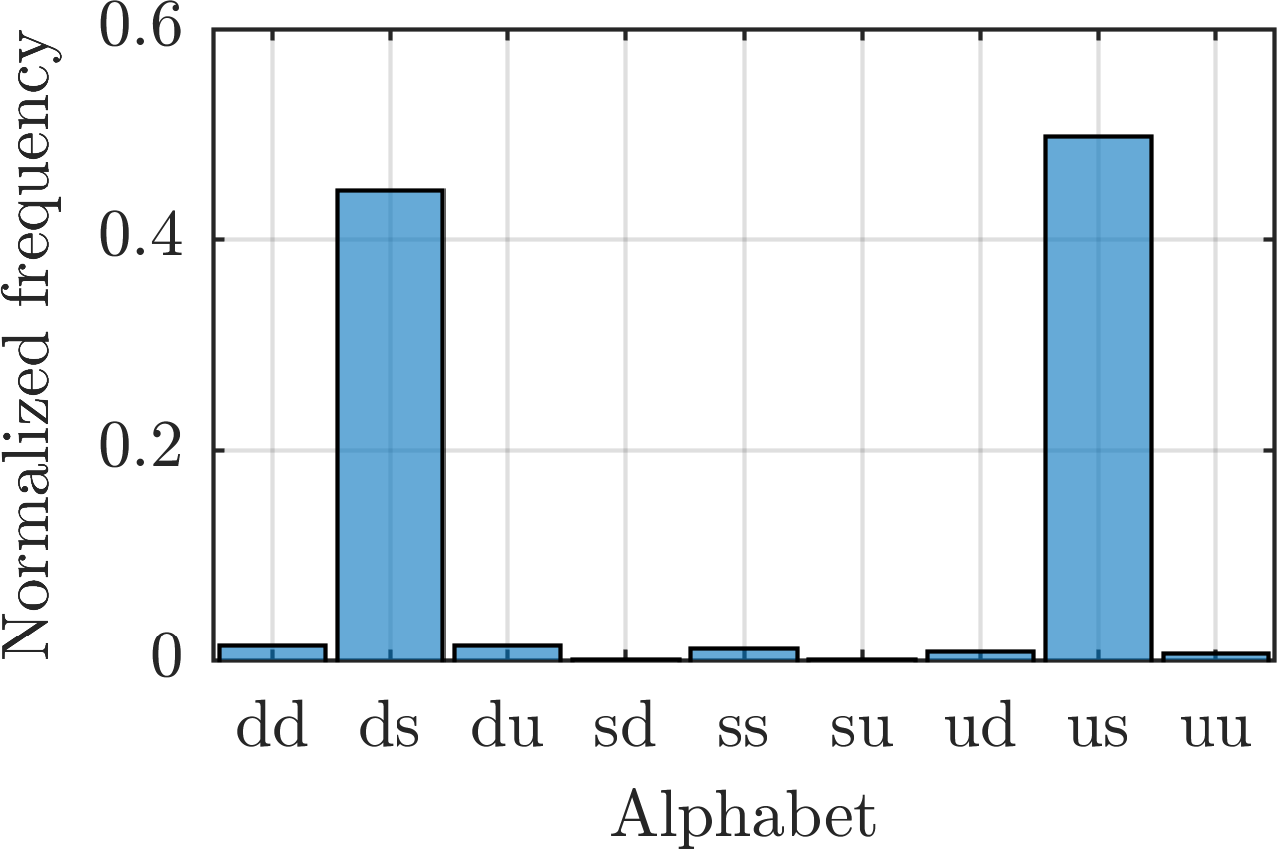}}
	\end{subfigure}
	\hfill
	\begin{subfigure}[Operation mode 2 - snippet]{
			\includegraphics[scale=1]{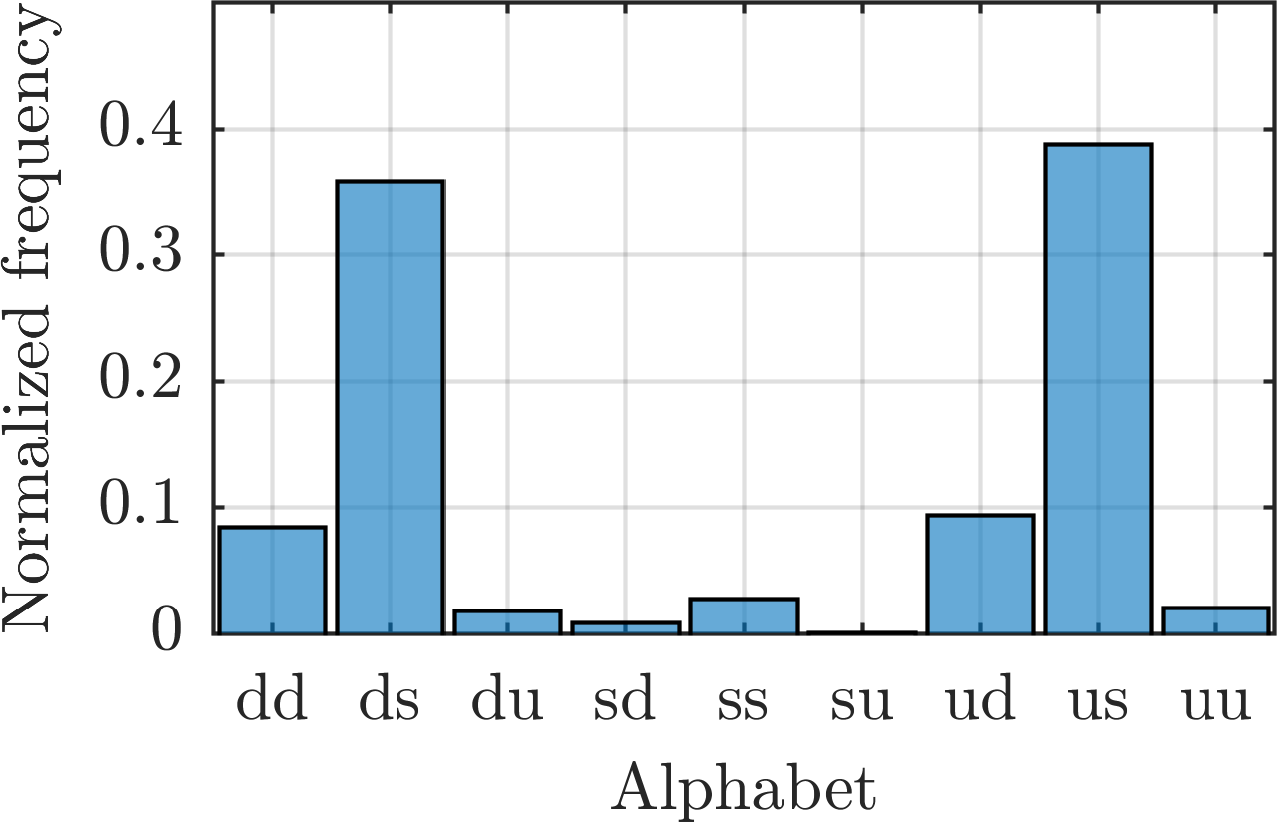}}
	\end{subfigure}
	\caption{Histograms of occurring symbol combinations of a machine in two different operation modes. Top: Histograms for the entire time range shown in Fig~\ref{fig:MCLA1} (top) and Fig~\ref{fig:MCLA2} (top); Bottom: Histograms for the signal snippets presented in Fig~\ref{fig:MCLA1} (bottom) and Fig~\ref{fig:MCLA2} (bottom).}
	\label{fig:freqdict}
\end{figure}

The generated symbolic representation is used for further analyses.
Building up histograms for occurring symbol combinations offers an insight in the overall behaviour of the system, see Fig.~\ref{fig:freqdict}).
This allows inter-machine comparison and comparison of different signal portions/ranges as well as classification of the operation mode.
On top of Fig.~\ref{fig:freqdict} the histograms of the entire signal ranges shown in Fig~\ref{fig:MCLA1} (top) and Fig~\ref{fig:MCLA2} (top) are presented.
The histograms for the typical repeating snippets, shown in Fig~\ref{fig:MCLA1} (bottom) and Fig~\ref{fig:MCLA2} (bottom), are visualized on the bottom.
Since the machine is interrupted several times in both operating modes, the bins for the stationary state (\texttt{ss}) are more visible for the entire signal sequences (top).
Excluding these bins, the statistics (histograms) of the shown snippets can act as representatives (motifs) for the operating modes.
It can be seen that the histograms differ whether the machine is operating in mode 1 (left) or mode 2 (right). Especially the occurrences of \texttt{dd} and \texttt{ud} reveal the differences. In future investigations the definition of a similarity measure for such histograms is planned to compare them qualitatively and may use this for automatic operation recognition and finding motifs.
Note: sorting the histograms in decreasing order of occurrences will yield a classical frequency dictionary.

A big advantage of the presented symbolic time series analysis is, that he sequence of symbols - either single or multi channel - can now be addressed with techniques more common to computational linguistics (\eg \emph{regex}) \cite{Clark2010ComputationalLinguistics}, which is a growing field of research.
%
%Some topics: frequency dictionary, modes of histogram, data compression, segmenting, histogram
	\section{Conclusion}
Successful data analytics in large physical systems must embed the modelling of the individual component and complete system dynamics.
This has been addressed by providing for a linear differential operator or its inverse in each and every signal- or derived-data-channel.
A multi-variate symbolic time series analysis has been introduced. It permits a symbolic view of the system and its dynamics.
The concept of frequency dictionaries has been applied to automatic operation recognition; this functions for operation types which are characterised by a specific distribution of symbols.
A major advantage of the proposed method is its intrinsic multi-scale property.
This enables the identification of very short events in very large data sets.
Currently, we are performing research on the relationships between the sequences of symbols and the metaphor of language.
Initial results indicate that this opens the door to take advantage of new methods emerging in computational linguistics.
	%
	%show linewitdh
	%\printinunitsof{cm}\prntlen{\textwidth}
	%\printinunitsof{cm}\prntlen{\linewidth}
	%
	%\input{Sec_Statistics}
	%\input{Sec_ApplicationsConclusions}
	%
	%
	%
	%% Bibliography
	%\input{bibliography}
	\bibliography{LiteraturPublicatons-2017-ITISE2017}
	\bibliographystyle{splncs}
\end{document}